\documentclass[showpacs,amsmath,amssymb,twocolumn, pra, floatfix]{revtex4}
\usepackage[dvips]{graphicx}
\input{epsf}

\begin{document}

\title{Entanglement of localized states}

\author{O. Giraud, J. Martin and B. Georgeot}
\affiliation{Laboratoire de Physique Th\'eorique,
Universit\'e Toulouse III, CNRS, 31062 Toulouse, France}

\date{October 25, 2007}

\begin{abstract}
We derive exact expressions for the mean value
of Meyer-Wallach entanglement $Q$ for localized random vectors
drawn from various ensembles corresponding to different physical situations.
For vectors localized on a randomly chosen subset of the basis,
$\langle Q \rangle$ tends for large system sizes
to a constant which depends
on the participation ratio,
whereas for vectors localized on adjacent basis states
it goes to zero as a constant over the number of qubits.  Applications
to many-body systems and Anderson localization are discussed.
\end{abstract}
\pacs{03.67.Mn, 03.67.Lx, 05.45.Mt}
\maketitle


Random quantum states have recently attracted a
lot of interest due to their relevance to the field
of quantum information.
Since they are useful
in various quantum protocols \cite{random}, efficient
generation of random and pseudo-random  vectors
\cite{emerson}
and computation
of their entanglement properties
\cite{ranvec} have been widely discussed.

Random states are not necessarily uniformly spread over the whole
Hilbert space. It is therefore natural to
study entanglement properties of
random states which are restricted to a certain subspace of Hilbert
space, or whose
weight is mainly concentrated
on such a subspace. Such states can appear naturally as
 part of a quantum algorithm, or can be imposed
by the physical implementation of qubits,
through e.\ g.\ the presence of symmetries.

In addition, random states built from Random Matrix Theory (RMT) have been
shown to describe many properties of complex quantum
states of physical systems, especially in a regime of quantum chaos.
Yet in many cases physical systems display wavefunctions which
are localized preferentially
on part of the Hilbert space.  This happens for example if there is a
symmetry, or when the presence of an
interaction delocalizes independent-particle states inside an
energy band given by the Fermi Golden Rule.  A different case
concerns Anderson localization of electrons, a much studied phenomenon where
wavefunctions of electrons in a random potential are exponentially localized.
  Assessing the
entanglement properties of such states not only enables
to relate the entanglement to other physical properties, but also
has a direct relationship with the algorithmic complexity of the
simulation of such states.  Indeed, it has been shown \cite{josza}
that weakly entangled states can be efficiently simulated on classical
computers.

For a vector $\Psi $ in a $N$-dimensional Hilbert space,
localization
can be quantified through the
Inverse Participation Ratio (IPR)
$\xi=\sum_i|\Psi_i|^2/\sum_i|\Psi_i|^4$ where
$\Psi_i$ are the components of $\Psi$.
This measure gives $\xi=1$ for a basis vector, and $\xi=M$
for a vector uniformly spread on $M$ basis vectors.

To investigate entanglement properties of localized vectors,
we choose the measure of entanglement
proposed in
\cite{MW}.  Meyer-Wallach entanglement (MWE) $Q$ can
be seen as an average measure of the bipartite entanglement
(measured by the purity) of one qubit with all others.
The quantity $Q$ has been widely
used as a measure of the entangling power of quantum maps \cite{chaosent},
or to measure entanglement generation in pseudo-random
operators \cite{emerson}. For a  pure
$N$-dimensional state $\Psi$ coded on $n$ qubits ($N=2^{n}$),
 $Q=2\left(1-\frac{1}{n}\sum_{r=0}^{n-1}R_r\right)$, where
$R_r=\textrm{tr}\rho_r^2$ is the purity of the $r$-th qubit
($\rho_r$ is the partial trace of the density matrix
over all qubits but qubit $r$). It
can be rewritten as $Q=\frac{4}{n}
\sum_{r=0}^{n-1}G(u^{r},v^r)$, where
$G(u,v)=\langle u|u\rangle\langle v|v\rangle-|\langle u|v\rangle|^2$
is the Gram determinant of $u$ and $v$, and $u^r$ (resp.\ $v^r$) is
the vector of length $N/2$ whose components are the $\Psi_i$ such
that $i$ has no (resp.\ has a) term $2^r$ in its binary
decomposition. Vectors $u^r$ and $v^r$ are therefore a
partition of vector $\Psi$ in two subvectors according to the value
of the $r$-th bit of the index.

Analytical computations will be made on ensembles of random
vectors.  In this case,
individual quantum states in a given basis have components
whose amplitudes, phases and positions in the basis are drawn from
a distribution according to some probability law.  Quantities
such as IPR or entanglement measure are then averaged over all
realizations of the vector.
A simple example of a random vector localized on $M$ basis states
can be constructed by taking $M$ components with equal
amplitudes and
uniformly distributed random phases, and setting all the others to
zero.
A more refined example consists in
using, as nonzero components, column vectors of $M\times M$ random
unitary matrices
drawn from the Circular Unitary Ensemble
of random matrices (CUE vectors).

In the first part of this paper we study entanglement properties of
random quantum states which are localized, or mainly
localized, in some subset of the basis vectors.  We show
that very different behaviors can be obtained depending on the
precise type of localization discussed.
The first case we consider (section \ref{section1})
consists in random states whose non zero components
in a given basis are randomly distributed among the basis vectors.
Moreover, these nonzero components are chosen to have random values.
Averages over random realizations therefore imply
that we average both over position of the nonzero components among the basis
vectors and over the random values of these nonzero components.
We show that the mean entanglement
can be
expressed as a function of the number of nonzero components of the vector.
We then
show that this result can be generalized. Indeed for any vector
with random values distributed according to some probability distribution,
the mean entanglement can in fact be expressed as a function of the mean IPR.
Notably, this function tends to a constant close to 1 for large system size.
While the vectors in section \ref{section1} are localized on computational
basis states which are taken at random, in section \ref{section2}
random vectors are localized on computational basis states which are
adjacent when the basis vectors are ordered according to the number which
labels them. In this case the mean entanglement can again be expressed as
a function of the mean IPR, but in contrast this function tends to 0 for large
system size. Again, the averages are performed both on position and values of
the components.
In the second part of the paper, we compare these
results to the entanglement of various physical systems which display
localization (section \ref{section3}).

The question of entanglement properties of localized states has already
been addressed in other works. The concurrence of certain localized states
in quantum maps has been studied in \cite{BetShe03,montangero}, but with
an emphasis on the effect of noise in quantum algorithms. In \cite{chinois},
a relation between the linear entropy and the IPR has been
derived in the special case where each qubit is an Anderson localized
state. During the course of this work, a
preprint appeared which uses different techniques to relate
the entanglement to the IPR \cite{viola} in the case of vectors
localized on non-adjacent basis states, as in section \ref{section1}.
Interestingly enough,
the formulas obtained in \cite{viola} are fairly
general. They are derived by different techniques and rest
on different assumptions.
In particular,
the authors of \cite{viola} do not average over random phases. They obtain a formula where
entanglement is expressed as a function of the mean IPR calculated in three different bases,
a quantity that is often delicate to evaluate.
Our work uses different techniques and the additional assumption
of random phases to
get a different formula (formula (3)) which involves only the IPR in one basis,
a quantity that can be easily evaluated in many cases and is directly
related to physical quantities such as the localization length.  For example,
it enables us to compute readily the entanglement for localized CUE vectors
(see (4)).
However there are instances of systems (e.g.\ spin systems) where these different formulas
give the same results.

\section{Analytical results for randomly distributed localized vectors}
\label{section1}
Let us first consider
a random state $\Psi$ of length $N=2^n$
in the basis $\{|i\rangle=|i_0\rangle\otimes\cdots\otimes|i_{n-1}\rangle,
0 \leq i \leq 2^{n}-1, i=\sum_{r=0}^{n-1}i_r 2^r\}$ of
register states (where all $ \sigma_{r}^z$ are diagonal).
Suppose the state $\Psi$ has $M$ nonzero
components which we denote
by $\psi_{i}$, $1\leq i\leq M$. Each nonzero component is random and
additionally
corresponds to a randomly chosen position among basis vectors.
The corresponding average will be denoted by $\langle ... \rangle$.
We make the assumption
that these components have
uncorrelated random phases,
and that $\langle |\psi_p|^2 \rangle$ and
$\langle |\psi_p|^2|\psi_q|^2 \rangle$ do not depend on
$p,q$.
We calculate the contribution to MWE
of a partition $(u,v)$ (we drop indices $r$).
Suppose $u$  has $k$ non-zero
components $u_i$, $i\in I$ and that $v$ has
$M-k$ non-zero components $v_j$, $j \in J$,
with $I,J$ subsets of $\{1,...,N/2\}$. We
define  $T=I \cap J$ and
the bijections $\sigma$ and $\tau$ such that $u_i=\psi_{\sigma(i)}$
and $v_j=\psi_{\tau(j)}$.
Setting $s_p=|\psi_p|^2$, the average $G(u,v)$ is given
by
\begin{equation}\label{guvo}
\langle G(u,v)\rangle=\Big\langle\sum_{p\in
\sigma(I)}s_{p}\sum_{q\in
\tau(J)}s_{q}\Big\rangle-\Big\langle\sum_{i\in
T}s_{\sigma(i)}s_{\tau(i)}\Big\rangle,
\end{equation}
where the non-diagonal terms in $|\langle
u|v\rangle|^2$ have vanished by integration over the random phases
of the $\psi_p$. We assumed that $\langle s_p s_q\rangle$ ($p\ne q$)
does not depend on $p,q$, thus
 $\langle
G(u,v)\rangle=[k(M-k)-t]\langle s_p s_q \rangle$, the overlap
$t$ being the number of elements of $T$.
Since $\langle
u|u\rangle+\langle v|v\rangle=1$, we also have $\langle G(u,v)\rangle
=k(\langle s_p\rangle-\langle s_p^2\rangle)-[k(k-1)+t]\langle
s_ps_q\rangle$. We then equate both expressions
and use our
hypothesis that $\langle |\psi_p|^2 \rangle$ and $\langle |\psi_p|^4
\rangle$ are independent of $p$, which implies that $\langle s_p \rangle =
1/M$ and $\langle s_p^2\rangle = \langle 1/\xi\rangle/M$, to get
\begin{equation}
\label{guvkt}
\langle G(u,v)\rangle=\frac{k(M-k)-t}{M(M-1)}\left(1-
\langle\frac{1}{\xi}\rangle\right).
\end{equation}
As this result depends only on
$(k,t)$, the calculation of $\langle
Q\rangle$ comes down to counting the number of
positions of the non-zero components in vectors $u$ and $v$
yielding the same pair $(k,t)$.
The combinatorial weight associated to a given
$(k,t)$
is $\binom{N/2}{k}\binom{k}{t}\binom{N/2-k}{M-k-t}$.  At fixed
$k$, $t$ ranges from $0$ to $\min(k,M-k)$. Summing
all contributions yields:
\begin{equation}\label{Qnadj}
\langle Q\rangle=\frac{N-2}{N-1}\left(1-\langle\frac{1}{\xi}\rangle\right).
\end{equation}

This result does not depend on $M$. It can in
fact be derived by an alternative method with
less restrictive assumptions. Let us sum up
all the localization properties of $\Psi$ in
the IPR $\xi$ alone, whatever the value of $M$.
We define the correlators
$C_{xx}=(\overline{ |u_i|^2 |u_j|^2  +|v_i|^2 |v_j|^2 })/2$,
and
$C_{xy}=\overline{ |u_i|^2 |v_j|^2 }$, where the overline denotes
the average  taken over all $n$ partitions $(u^r,v^r)$ corresponding to the
$n$ qubits, and over all $i,j \in \{1,...,N/2\}$ with $i\neq j$ (for $C_{xx}$)
and all $i,j \in \{1,...,N/2\}$ (for $C_{xy}$). Thus $C_{xx}$ quantifies
the internal correlations inside $u$ and $v$, and $C_{xy}$
the cross correlations between $u$ and $v$.
Normalization imposes that $\langle1/\xi\rangle + N(N/2-1)\langle
C_{xx}\rangle +(N^2/2) \langle C_{xy}\rangle =1$,
and Eq.~(\ref{guvo}) leads to
$\langle Q \rangle= N(N-2)\langle C_{xy}\rangle $.  The assumption $\langle
C_{xx}\rangle= \langle C_{xy}\rangle$
is then sufficient to get Eq.~\eqref{Qnadj}.
This derivation also shows that
if the phases are uncorrelated and formula  (\ref{Qnadj}) does not apply,
then necessarily $\langle C_{xx}\rangle \neq\langle C_{xy} \rangle$.

Our result Eq.~(\ref{Qnadj}) involves only the mean IPR in one basis,
and uses the assumptions that on average cross correlations are equal
to internal correlations for the partitions, whatever the probability
distribution of the components, and that random phases are uncorrelated.
This is to be compared with the result in \cite{viola} where
$\langle Q \rangle$
is related to the sum of IPR for three mutually unbiased bases. Their result
does not use the assumption of uncorrelated random phases, but requires
a stronger hypothesis on correlations
(namely, that vector component correlations in average do not
depend on the Hamming distance between the corresponding vector
component indices).
In particular, our formula (\ref{Qnadj}) allows to compute
$\langle Q \rangle$ e.\ g.\
for a CUE vector localized on $M$ basis vectors;
in this case  $\xi=(M+1)/2$, and
we get
\begin{equation}
\label{QnadjCUE}
\langle Q\rangle=\frac{M-1}{M+1}\frac{N-2}{N-1}.
\end{equation}
In \cite{Lub}, Lubkin derived an expression for the mean MWE for
non-localized CUE vectors of length $N$, giving
$\langle Q\rangle=(N-2)/(N+1)$. Consistently, our formula yields
the same result if we take $M=N$. For a
vector with constant amplitudes and random phases on $M$ basis vectors,
$\xi=M$ and
\begin{equation}
\langle Q\rangle=\frac{M-1}{M}\frac{N-2}{N-1}.
\end{equation}
Formula (\ref{Qnadj}) can be easily modified to account for
the presence of symmetries. For instance, suppose the system
presents a symmetry which does not mix basis
states within two separate subspaces of dimension $N/2$.
It is then easy to check that $N$ in (\ref{Qnadj}) should be replaced
by $N/2$.

\section{Analytical results for adjacent localized vectors}
\label{section2}
Up to now we have considered random vectors whose components were
distributed over a randomly chosen subset of basis vectors. However
in many physical situations vectors are localized preferentially on
particular subspaces of Hilbert space. An important case consists in
random vectors localized on $M$ computational basis states which are
adjacent when the basis vectors are ordered according to the number which
labels them. The general form of such a vector would be
$|c\rangle$,...,$|c+M-1\rangle$, $0\leq c\leq 2^n-1$.
Again, averaging over random realizations of the coefficients of $\Psi$
we get Eq.(\ref{guvkt}). The calculation of $\langle Q\rangle$
therefore reduces to determine $k$ and $t$ for all qubits and all
possible choices of the basis vectors on which $\Psi$ has nonzero
components.
For a given $r$, vectors $u$ and $v$ correspond to a partition
of the set of the components  $\Psi_i$ of $\Psi$ according
to the value of the $r$-th bit of $i$. For instance for the qubit
$r=1$, and $M=9, N=16$, a typical realization of vectors $u$ and $v$
would be
\begin{equation}
\begin{array}{ccccccccccc}
u ={}& ( &0 &0 & 0 &\psi_1 &\psi_4 &\psi_5 &\psi_8& \psi_9 &),\\
v ={}& ( &0 &0 &\psi_2 &\psi_3& \psi_6 &\psi_7& 0& 0 &).
\end{array}
\end{equation}
Each vector $u$ and $v$ can be split into $2^{n-1-r}$ blocks of
length $2^r$. There are $N n$ ways of constructing such pairs
$(u,v)$, by choosing a qubit $r$ and a position $c$ for $\psi_1$.
The numbers $k$ and $t$ depend on three quantities: the label
$r\in\{0,\ldots,n-1\}$ of the qubit whose contribution is
considered; the position $c_r\in\{0,\ldots,2^r-1\}$
 of $\psi_1$ within
a block, either in $u$ or in $v$;
the remainder $m_r$ of $M$ mod $2^{r+1}$.
Let $r_0$ be such that $2^{r_0-1}< M\leq2^{r_0}$. One has to
distinguish the contributions coming from qubits such that $0\leq
r<r_0$ and qubits such that $r\geq r_0$. First consider
 $0\leq r<r_0$. Suppose $\psi_1$ is a component of vector $u$.
 One can
check that $I\cup J$ has $k+t+c_r=M$ elements, and
$ I\setminus T$ has
 $k-t=g_r(m_r+c_r)$ elements, where
$g_r(x)=2^rg(x/2^r)$ with $g(x)=|1-|1-x||$, $x\in[0,3[$. These two
equations lead to $k=\frac{1}{2}(M-c_r+g_r(m_r+c_r))$ and
$t=\frac{1}{2}(M-c_r-g_r(m_r+c_r))$. Similarly, when $\psi_1$ is a
component of vector $v$, we get
$k=\frac{1}{2}(M+c_r-g_r(m_r+c_r))$ and
$t=\frac{1}{2}(M+c_r-2^{r+1}+g_r(m_r+c_r))$. Altogether
this leads to $2 \times 2^r$ different contributions with
multiplicity $2^{n-1-r}$ (the number of blocks).
If $r\geq r_0$, $t$ is always zero and as
the position $c_r$ is varied, $k$ runs over $\{1,...,M-1\}$. Summing
all contributions together we get
\begin{eqnarray}
\label{Qmoy} \langle
Q\rangle&=&\left[\left(\frac{M-2}{M-1}r_0+\frac{2(2^{r_0}-1)}{M(M-1)}
+\frac{4}{3}\frac{(M+1)(2^n-2^{r_0})}{2^{n+r_0}}\right.\right.\nonumber\\
&-&\left.\left.\frac{1}{M(M-1)}\sum_{r=0}^{r_0-1}\chi_r(m_r)\right)
\left(1-\langle\frac{1}{\xi}\rangle\right)
\right]\frac{1}{n},
\end{eqnarray}
where $\chi_r(x)=\chi_r(2^{r+1}-x)=x^2-\frac{2}{3}x(x^2-1)/2^r$ for
$0\leq x\leq 2^r$. Equation~\eqref{Qmoy} is an exact formula for
$M\leq N/2$. For fixed $M$ and $n\rightarrow \infty$,
$n\langle Q\rangle$ converges to a constant $C$
which is a function of $M$ and $\xi$.
For $M=2^{r_0}$, $r_0<n$, all remainders
$m_r$, $r<r_0$ are zero, and Eq.~\eqref{Qmoy} simplifies to
\begin{eqnarray}
\label{Qmoyapprox} \langle
Q\rangle&=&\left[\left(\frac{(r_0+\frac{4}{3})M^2-2(r_0-1)M-\frac{10}{3}}{M(M-1)}\right.\right.\nonumber\\
&-&\left.\left.\frac{4(M+1)}{3
N}\right)\left(1-\langle\frac{1}{\xi}\rangle\right)\right] \frac{1}{n}.
\end{eqnarray}
Numerically, this expression with $r_0=\log_2M$
gives a very good approximation to
Eq.~\eqref{Qmoy} for all $M$.

Equation~\eqref{Qmoy} is exact for e.\ g.\ uniform and CUE vectors, and
we will see in section \ref{section3} that it
can be
applied even when $\Psi$ is not strictly zero outside a $M$-dimensional
subspace.

\section{Application to physical systems}
\label{section3}
We now turn to the application of these results
to physical systems.
Localized vectors randomly
distributed over the basis states
may model eigenstates of a many-body Hamiltonian with
disorder and interaction. Indeed, the latter
generically display a
delocalization in energy characterized by RMT statistics
of eigenvalues within a certain energy range, whereas
the distribution of eigenvector components
is Lorentzian or Gaussian.
As an example we choose the system governed by the
Hamiltonian
\begin{equation}
\label{hamil}
H = \sum_{i} \Gamma_i \sigma_{i}^z + \sum_{i<j} J_{ij}
\sigma_{i}^x \sigma_{j}^x.
\end{equation}
This model was introduced
in \cite{qchaos} to describe a quantum computer in presence of static
disorder.  Here
the $\sigma_{i}$ are the Pauli matrices for the qubit $i$.
The energy spacing between the two states of qubit $i$ is given
by $2\Gamma_i$ randomly and uniformly distributed in the interval
$[\Delta_0 -\delta /2, \Delta_0 + \delta /2 ]$.
The couplings $J_{ij}$ represent a random static interaction
between qubits and are
uniformly distributed in the interval $[-J,J]$.
For increasing
interaction strength $J$
eigenstates are more and more delocalized in the basis of
register states, and
a transition towards a regime of quantum chaos
takes place, with eigenvalues statistics close to the ones of RMT
\cite{qchaos}.
In parallel, this process leads to an increase of
bipartite entanglement in the system
\cite{italians}.  In the following, results will
be averaged over random realizations of the $\Gamma_i$ and $J_{ij}$
in \eqref{hamil} (``disorder realizations''),
which will be denoted by $\langle ...\rangle$.

The Hamiltonian \eqref{hamil}
presents a symmetry which does not mix basis states having even and odd
number of qubits in the state $|1\rangle$. Each symmetric subspace
contains $N/2$ basis vectors among which for each qubit $N/4$ have
value $|1\rangle$ and $N/4$ have value $|0\rangle$.
In this case, as explained at the end of Section I, $N$ has
to be replaced by $N/2$ in \eqref{Qnadj}. This symmetry has the
additional effect of making the second term
in (\ref{guvo}) vanish identically for all eigenvectors.

\begin{figure}[hbt]
\begin{center}
\includegraphics[width=.95\linewidth]{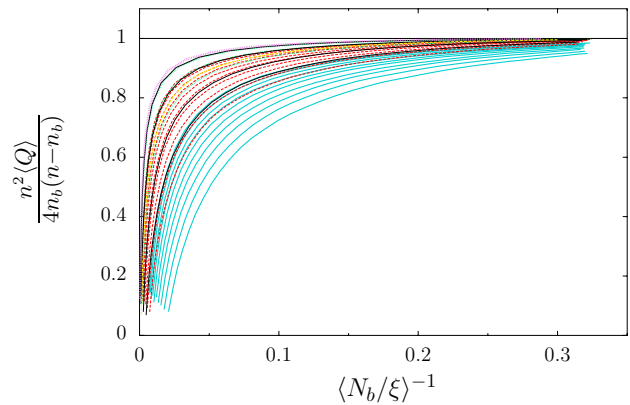}
\end{center}
\caption{(Color online) Scaled entanglement with respect to the
reduced localization length $\langle N_b/\xi\rangle^{-1}$
with $N_b=\binom{n}{n_b}$. The
blue thick solid curves correspond to the second band (for $n=11$--$20$), the
red thin dashed curves to the third band (for $n=11$--$20$), the green
thin dotted curves to the fourth band (for $n=13,15,17$), the black thin solid
curves to the fifth band (for $n=10,11,13,15$), the yellow thick dashed
curve to the sixth band (for $n=12$), and the violet thick dotted curve to the
seventh band (for $n=14$).} \label{bands}
\end{figure}

Before applying the results of section \ref{section1}
to the more generic case $\delta \approx \Delta_0$,
we first briefly discuss the specific case $\delta \ll \Delta_0$.
In this case, the energy
spectrum of the system
is divided into bands corresponding to register states with the same number
$n_b$ of qubits in the $|1\rangle$ state.
Delocalization takes place inside each band separately,
corresponding to a reduced Hilbert space of dimension
$N_b=\binom{n}{n_b}$.
In this case, all the basis states
on which the delocalization takes place have $n_b$ qubits among $n$
in the state $|1\rangle$. This implies that the components
of the wave function are not symmetrically distributed
on the two vectors $u$ and $v$ of Eq.\eqref{guvo}.
Thus, the correlation assumption breaks down and Formula (\ref{Qnadj})
does not
apply. However we can derive a specific formula in this case, starting
back from  Eq.\eqref{guvo}.
The probabilities that a basis vector with $n_b$ qubits in $|1\rangle$
enters into $v$ and $u$ are respectively $\eta=n_b/n$ and $1-\eta$.
So we
expect the norm of $u$ to be on average $1-\eta$,
and the norm of $v$ to be on average $\eta$. This
implies that for a homogeneously delocalized vector one has
$\langle Q \rangle \rightarrow 4 \eta (1-\eta)$ for $n\rightarrow \infty$
and $\eta$ constant, since the presence of the symmetry
makes the second term in \eqref{guvo} vanish.
Applying this latter formula to the specific case $n_b=1$, we recover
the result derived in \cite{chinois}.
Thus  $\langle Q\rangle$ tends to a value between $0$ and $1$
depending on the band, as can be seen
numerically in Fig.~\ref{bands}.
\begin{figure}[hbt]
\begin{center}
\includegraphics[width=.95\linewidth]{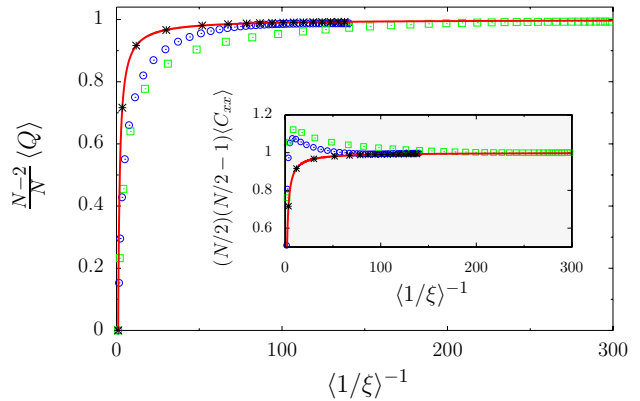}
\end{center}
\caption{(Color online) Scaled mean MWE $\langle Q \rangle
(N-2)/N$ of \eqref{hamil} vs IPR
for $\delta =\Delta_0$, $n=10$ (blue circles)
and $n=11$ (green squares). Average is over
$N/16$ central eigenstates and $100-200$ disorder realizations.
Red solid line is the theory, and stars are the data for
$n=10$ with random shuffling of components.
Inset: scaled correlator $(N/2)(N/2-1)\langle C_{xx} \rangle$ with
same parameters; red line is the result when $\langle C_{xx}\rangle=\langle
C_{xy}\rangle$.} \label{xi}
\end{figure}

In the case where $\delta \approx \Delta_0$, the bands become mixed by the
interaction, and delocalization takes place inside the whole
Hilbert space.
Formula (\ref{Qnadj}) should apply, once modified to take
into account the symmetry of the Hamiltonian \eqref{hamil}.
A straightforward modification of the reasoning leading to Eq.~\eqref{Qnadj}
yields $\langle Q\rangle=\frac{N}{N-2}\left(1-\langle\frac{1}{\xi}\rangle
\right)$. It turns out that the presence of this particular
symmetry allowed the authors of \cite{viola} to make explicit
their formula in a similar case, yielding the same expression
as ours.

Figure~\ref{xi} shows the entanglement of eigenvectors of Hamiltonian
\eqref{hamil}
compared to this formula.  The entanglement goes to one, but departs
from the formula at some values of the IPR $\xi$.  The inset illustrates
 that this discrepancy corresponds to a breakdown of the hypothesis
 $\langle C_{xx}\rangle =\langle C_{xy}\rangle$, because of correlations.  These correlations
are probably due to the perturbative regime where delocalization takes
place on a strongly correlated subset of states.
Figure~\ref{xi} shows that if these correlations
are destroyed by random permutations of the
components, the results
are in perfect agreement with the theory, eventhough the distribution
of the component amplitudes is left unchanged.
This confirms that (\ref{Qnadj}) can be applied if correlations
are weak between the vector components, whatever their distribution.

\begin{figure}
\begin{center}
\includegraphics[width=.95\linewidth]{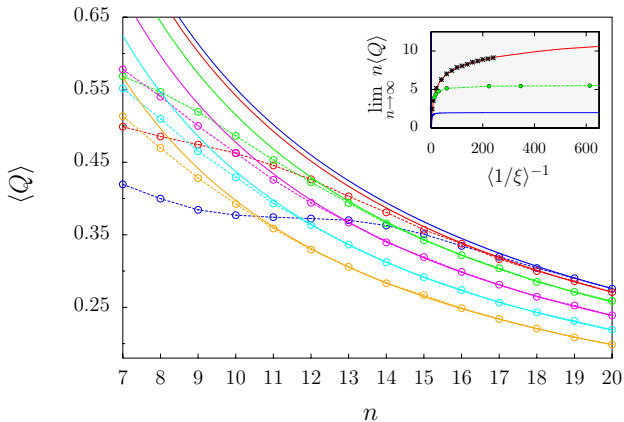}
\end{center}
\caption{(Color online) Mean MWE vs the number of qubits for 1D
Anderson model with disorder from top to bottom $w=0.2$ (blue),
$0.5$ (red), $1.0$ (green), $1.5$ (magenta), $2.0$ (cyan), and $2.5$
(orange). Average is over 10 central eigenstates for 1000 disorder
realizations. Solid lines are the $C/n$ fits of the tails. Inset:
Value of $C=\lim_{n\to\infty}n\langle Q\rangle$ as a function of IPR
$\xi$ (green dots) for the values of $w$ above and $w=0.4$, together
with analytical result from Eq.~\eqref{Qmoy} (red line, top) and
from Eq.~\eqref{Qmarre} (blue line, bottom) both for $M=2\xi$ .
Stars are the $C$ values resulting from a $C/n$ fit of the numerical
data for CUE vectors of size $N$ with exponential envelope
$\exp(-x/l)$.} \label{Qanderson}
\end{figure}

In the case of localization on adjacent basis vectors,
formula (\ref{Qmoy}) can be compared to wavefunctions
of electrons in the regime of Anderson localization.
Indeed, one dimensional disordered Anderson model is known to display
localized eigenstates for any strength of disorder. This type of
localization is a one-body phenomenon, but it has been shown that
it can be efficiently simulated on a $n$-qubit quantum computer,
$\Psi$ describing the particle in the position representation
\cite{pomeransky}.
The localization of the particle takes place
on a certain number of adjacent computational basis vectors,
and the entanglement of the quantum state is related to the
entanglement produced by the quantum algorithm.
The wavefunctions of the system are known to have an envelope
of the form $\exp(-x/l)$ where $l$ is the localization length.
For $N$-dimensional CUE vectors
with such an exponential envelope, we checked that
$\langle Q\rangle$ is in excellent agreement
with \eqref{Qmoy} with $\xi=l$ and $M=2\xi$
(stars in inset of Fig.~\ref{Qanderson}).
To test the formula on actual wavefunctions of the Anderson model,
we consider a
one dimensional chain of vertices with nearest-neighbor coupling
and randomly distributed on-site disorder, described by the
Hamiltonian $H_0+V$. Here $H_0$ is a diagonal operator whose elements
$\epsilon_i$ are Gaussian random variables with variance $w^2$, and
$V$ is a tridiagonal matrix with non-zero elements only on the first
diagonals, equal to the coupling strength, set to $1$.
For this system, $\langle ... \rangle$ therefore means averaging over
the diagonal random values.
Figure~\ref{Qanderson} displays $\langle
Q\rangle$ calculated numerically for eigenvectors of
this system, as a function of the number $n$ of qubits for various
strengths of the disorder $w$.
 The expected decrease as $C/n$ is
perfectly reproduced for large enough values of $n$. The inset shows
the value of the constant $C$ compared to the theory
\eqref{Qmoy}, as a function of  $\xi$. The deviation from
\eqref{Qmoy}, in particular the saturation for large $\xi$,
 can be understood by looking at the structure of
eigenvectors in Anderson model: when there is no disorder ($w=0$)
the eigenvalues are $E_k=2\cos 2\pi\nu_k$ and eigenvectors are
plane waves with frequency $\nu_k$. For weak disorder eigenvectors
are exponentially localized with localization length $\xi$ but still
oscillate at frequencies distributed as a Lorentzian of width
$1/\xi$ around $\nu_k$. We chose eigenvectors with energy $E_k \approx 0$
($\nu_k \approx 1/4$),
yielding rapid oscillations of period 4 which strongly decrease
entanglement. It is easy to adapt the analysis leading to
Eq.~\eqref{Qmoy} for $\Psi$ chosen as e.\ g.\ a vector
with $\Psi_j=\cos\pi j/2$, $c+1\leq j\leq c+M$, and zero elsewhere.
For instance for
$M=2^{r_0}$, $r_0<n$, we get (averaging over $c$)
\begin{equation}
\langle Q\rangle=
\Big(\frac{26}{9}-\frac{4}{M}-\frac{8(3r_0+1)}{9M^2}
-\frac{4(M^2-4)}{3M 2^n}\Big)\frac{1}{n}. \label{Qmarre}
\end{equation}
Asymptotically $n\langle Q\rangle$ converges to a constant
independent of $\xi=M/2$.
The inset of Fig.~\ref{Qanderson} shows that this theory
captures the behavior of the numerical $\langle
Q\rangle$, although the saturation constant is different.
\begin{figure}[hbt]
\begin{center}
\includegraphics[width=.95\linewidth]{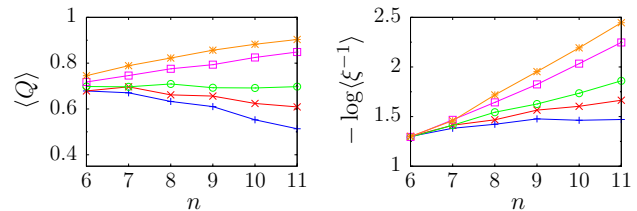}
\end{center}
\caption{(Color online) Mean MWE (left) and IPR
(right) vs number of qubits for
quantum smallworld networks with
$w=1$
and $p=0.001$ (blue $+$), $0.005$ (red $\times$), $0.01$
(green $\circ$), $0.03$ (magenta $\square$), $0.06$ (orange $\star$).
Logarithm is decimal.} \label{transition}
\end{figure}

Let us now add to this system $pN$ links between
randomly chosen vertices.  This additional
long-range interaction between few vertices turns the system into a quantum
smallworld network.
Such systems can be efficiently simulated on a quantum computer,
and display a localization-delocalization transition for fixed $w$ when
$p$ is increased \cite{smallworld}.
Figure~\ref{transition} shows that
this transition can be probed
through the entanglement of the system.
Indeed,
for small $p$ all eigenstates are exponentially localized; $\langle Q\rangle$
is given by \eqref{Qmoy} and decreases asymptotically
as $1/n$; when $p$ is increased the delocalization transition takes
place and $\langle Q\rangle$ is now given by Eq.~\eqref{Qnadj}: for large $n$,
it saturates at $1-\langle 1/\xi\rangle$.

In conclusion, we have shown that in localized random states the mean MWE
can be directly related to the IPR $\xi$.
Entanglement properties are very different if the localization is
on adjacent basis vectors or not.
  Comparison with physical systems show that
global entanglement properties are reproduced,
although some discrepancies show that they
are much more sensitive than e.g. level statistics to the
details of the system.

We thank K. Frahm for helpful discussions,
CalMiP and IDRIS
for access to their supercomputers, and
the French ANR
(project INFOSYSQQ) and the IST-FET program of the EC
(project EUROSQIP) for funding.

\end{document}